\input{aipcheck}

\documentclass[
    ,final            
  ]
  {aipproc}

\layoutstyle{6x9}

\begin{document}

\title{Geodesic Completeness around Sudden Singularities}

\classification{04.20.Dw, 98.80.Jk}
\keywords      {Sudden singularities, Big Rip, Cosmology, Geodesics}

\author{L. Fern\'andez-Jambrina}{
  address={E.T.S.I. Navales, Universidad
Polit\'ecnica de Madrid,
Arco de la Victoria s/n,  E-28040 Madrid, Spain}, 
email={leonardo.fernandez@upm.es}
}

\author{R. Lazkoz}{
  address={F\'\i sica Te\'orica, Facultad de Ciencia y Tecnolog\'\i a, Universidad del
Pa\'\i s Vasco, Apdo.  644, E-48080 Bilbao, Spain}, 
email={ruth.lazkoz@ehu.es}
}

\begin{abstract}
    In this talk we analyze the effect of recently proposed classes of
    sudden future singularities on causal geodesics of FLRW
    spacetimes.  Geodesics are shown to be extendible and just the
    equations for geodesic deviation are singular, although tidal
    forces are not strong enough to produce a Big Rip.
\end{abstract}

\maketitle


\section{Introduction}

The experimental evidence of accelerated expansion of the universe
(supernovae type Ia, redshift of distant objects and fluctuations of
background radiation) implemented in cosmological ghost models,
$\rho+p<0$, where $\rho$ is the density of the matter content and $p$ 
is the pressure, 
leads to an unexpected kind of singularities, named as ``big rip'', since
they would destroy galactic structures and even atoms due to 
accelerated expansion \cite{caldwell} of the universe. 

In FLRW cosmologies these singularities are characterized by
an infinite scale factor at a finite proper time, $a(t_{s})\to
\infty$. 

These FLRW models make use of perfect fluid equations of state that
violate all energy conditions (weak, strong and dominant), since
$\rho+p<0$.
    
%
%
\section{Sudden singularities}
    
Barrow \cite{barrow} suggests that such behaviour may be attained even if the strong
energy condition is fulfilled
\[\rho>0\;,\qquad \rho+3p>0\;,\] by rejecting linear equations of
state, $p\not= w \rho$. We may see it by taking a look at Friedmann equations
for FRLW cosmologies with Hubble constant $H:={ a'}/{a}$,
    
\begin{eqnarray*}
&&3H^2=\rho-\frac{k}{a^2}\;,\\
&& \rho'+3H(\rho+p)=0\;,\\
&&\frac{ a''}{a}=-\frac{\rho+3p}{6}\;.
\end{eqnarray*}

If we require $a(t_{s})\;, H(t_{s})<\infty$ (that is, finite
$a(t_{s})\;,  a'(t_{s})$), we obtain finite density
$\rho(t_{s})<\infty$, but still a singularity may occur if $p(t)\to
\infty$ at $t_{s}$. 
Of course, this means $a''/{a}\to-\infty$.
 
Barrow achieves this behaviour with the following model, 
\begin{eqnarray*}
    &&ds^2=-dt^2+a(t)\left\{\frac{dr^2}{1-kr^2}+ r^2\left(d\theta^2+\sin^2\theta
    d\phi^2\right)\right\},\nonumber\\
    &&a(t)=1+
    \left(\frac{t}{t_{s}}\right)^q(a_{s}-1)-\left(1-\frac{t}{t_{s}}\right)^n,
\end{eqnarray*}
with constants $a_{s}=a(t_{s})$, $0<q\le 1$, $1<n<2$, $k=0,\pm1$.

Of course, since
\[
a''(t)=q(q-1)\frac{a_{s}-1}{t_{s}^q}t^{q-2}-\frac{n(n-1)}{t_{s}^2(1-t/t_{s})^{2-n}}
\to-\infty\;,\]
there is a 
Big Bang singularity at $t=0$ and a ``sudden'' singularity at $t=t_{s}$.

As we required, since $\rho$ and $p$ are positive, weak and strong energy conditions
are fulfilled. But $\rho-p$ is negative at some time. Therefore the dominant energy
condition is violated \cite{lake}. 
Further examples of this behaviour have been shown by Barrow
\cite{barrow1},
Dabrowski, for inhomogenous models \cite{dabrowski} and Nojiri and Odintsov, by
considering quantum corrections \cite{odintsov}. 
The sudden singularity is even ``milder'' if $a''$ is finite and higher
derivatives diverge for larger values of $n$. 

\section{Geodesic completeness}

These models therefore show a curvature singularity at $t_{s}$, though
both $a$ and $H$ are finite there. 
This behaviour is different from
Big Rip singularities found in phantom cosmologies.

However, the usual definition for singularities refers to incomplete
causal geodesics \cite{HE}.  It is reasonable to ask whether causal geodesics
are incomplete in Barrow's models.

Since FLRW spacetimes are homogeneous and isotropic, they have six
independent isometry generators,
    \begin{eqnarray*}\xi_{1}&=&\frac {\sin
    \theta \cos \phi }{f(r) }\,\partial_{r} +\frac {\cos\theta \cos \phi
    }{rf(r)}\,\partial_{\theta}-\frac {\sin \phi}{rf ( r) \sin \theta
    }\,\partial_{\phi},~~~\\\xi_{2}&=&\frac {\sin \theta \sin\phi }{f(r)
    }\,\partial_{r} +\frac {\cos\theta \sin \phi
    }{rf(r)}\,\partial_{\theta}+\frac {\cos \phi}{rf ( r) \sin \theta
    }\,\partial_{\phi},~~~\\\xi_{3}&=&\frac{\cos\theta}{f(r)}\,\partial_{r}-\frac{\sin\theta}{rf(r)}\,\partial_{\theta},~~~\\\zeta_{1}&=&\cos\phi\,\partial_{\theta}-\cot\theta
    \sin\phi\,\partial_{\phi},~~~\\\zeta_{2}&=&\sin\phi\,\partial_{\theta}+\cot\theta   
    \cos\phi\,\partial_{\phi},~~~\\\zeta_{3}&=&\,\partial_{\phi},~~~\end{eqnarray*}
which produce six constants of geodesic motion for a geodesic parametrized by its proper time
$\tau$,
{\small\begin{eqnarray*}
P_{1}&=&a(t)\left\{\frac {
r\left(\cos \theta\cos \phi \,\dot\theta-\sin \theta \sin \phi
\,\dot \phi\right)}{f(r)} +
f(r) \sin \theta \cos \phi \,\dot r\right\},\\
P_{2}&=&a(t) \left\{\frac {
r\left(\cos \theta\sin \phi\,\dot\theta+\sin \theta \cos \phi
\,\dot \phi\right)}{f(r)}+f(r) \sin \theta \sin \phi \,\dot r\right\},\\
P_{3}&=&a(t)\left( f(r) \cos \theta \,\dot r-{ r }\sin
\theta\,\dot\theta/{f(r)}\right),\\L_{1}&=&a(t) r^2\left(\cos \phi
\,\dot\theta-\sin \theta \cos\theta\sin \phi \,\dot\phi\right),\\
L_{2}&=&a(t) r^2\left(\sin \phi \,\dot\theta+ \sin \theta
\cos\theta\cos \phi \,\dot\phi\right),\\L_{3}&=&a(t)r^2 \sin^2
\theta\,\dot\phi\;.
\end{eqnarray*}

Geodesic differential equations for timelike ($\delta=1$) and lightlike ($\delta=0$)
reduce to first order,
    \begin{eqnarray*}\dot t^2&=&\delta +
    \frac{P^2+k L^2}{a(t)}\;,\\\dot r&=&\frac {P_{1}\,\sin \theta \cos \phi
    +P_{2}\,\sin \theta \sin \phi +P_{3}\,\cos\theta}{a(t)f(r)}\;,\\\dot
    \theta&=&\frac {L_{1}\cos\phi+L_{2}\sin \phi
    }{a(t)r^2}\;,\\\dot\phi&=&\frac {L_{3}}{a(t)r^2 \sin^2
    \theta}\;,\\
    P^2&=&P_{1}^2+P_{2}^2+P_{3}^2\;,\qquad
    L^2=L_{1}^2+L_{2}^2+L_{3}^2\;.
    \end{eqnarray*}
    
The system may be further simplified, since due to spherical symmetry
every geodesic may be fit in the hypersurface $\theta=\pi/2$, with
$L_{1}=L_{2}=0=P_{3}$, by a suitable choice of the coordinates, then
    \begin{eqnarray*}\dot t^2&=&\delta +
    \frac{P^2+k L^2}{a(t)}\;,\\\dot r&=&\frac {P_{1}\,\cos \phi +P_{2}\,
    \sin \phi}{a(t)f(r)}\;,\\\dot\phi&=&\frac
    {L_{3}}{a(t)r^2}\;,
    \end{eqnarray*}

These equations are not singular for finite $a(t)$. Therefore, we realize that  geodesics just see the Big Bang
singularity at $t=0$, but not the sudden singularity at $t=t_{s}$
\cite{ruth}. Furthermore, since $a$, $ a'$ are finite at $t_{s}$, the acceleration
vector of the geodesic, $(\ddot t, \ddot r, \ddot\theta, \ddot \phi)$,
which comprises the effect of inertial forces, is also regular. Only the third derivative is singular at $t_{s}$, but we just require
first and second derivatives to define geodesic equations.

\section{Tidal forces}    
    
Causal geodesics in such universes do not see the singularities but
through geodesic deviation effects, since they are due to the Riemann
tensor. Point particles travelling along causal geodesics do not experience
any singularity, but extended objects might suffer infinite tidal forces at
$t=t_{s}$.

According to Tipler's definition \cite{tipler} a strong curvature singularity is
encountered at a point $p$ if every volume element defined by three
linearly independent, vorticity-free, geodesic deviation vectors along
every causal geodesic through $p$ vanishes at this point.

Clarke and Krolak \cite{clarke} provide necessary and sufficient conditions for the
appearance of strong curvature singularities. 
If a causal geodesic meets a strong singularity at a value
    $\tau_{s}$, 
    \begin{eqnarray*}\int_{0}^{\tau}d\tau'\int_{0}^{\tau'}d\tau''|R^{i}_{\ 
	0j0}(\tau'')|\;,\end{eqnarray*}
    will diverge along the geodesic on approaching $\tau_{s}$. 
    
In the case of sudden singularities, the components of the Riemann
tensor diverge as $a''$, since $a'$ and $a$ are finite; and in the
worst case they diverge as a power $n-2$, for $1<n<2$.  Therefore
after one integration of the components of the Riemann tensor, the
power will be positive and the integral will not diverge. 
Hence sudden singularities are not strong according to Tipler
definition and therefore tidal forces do not crush all finite bodies. 
The spacetime may be extended across sudden singularities and cannot
be considered the final fate of these universes \cite{ruth}.

\section{Conclusions}

We have shown that causal geodesics are not affected by
sudden future singularities, since these singularities are not seen by
geodesic equations. 

Furthermore, considering just curvature singularities, it has been
shown that they are weak according to Tipler's 
definition and therefore finite objects are not necessarily torn on
crossing the singularities.  This is in contrast with Big Rip
singularities and therefore sudden singularities produce no Big Rip.

\begin{theacknowledgments}
    L. F.-J. is supported by the Spanish Ministry
    of Education and Science through research grant FIS2005-05198. R.L.
    is supported by the University of the Basque Country through
    research grant UPV00172.310-14456/2002 and by the Spanish Ministry
    of Education and Science through research grant FIS2004-01626.
\end{theacknowledgments}

\end{document}